\documentclass[a4paper,11pt]{article}

\usepackage{jinstpub} 
\usepackage{lineno}

\usepackage{graphicx}
\usepackage{booktabs}
\usepackage{siunitx}
\usepackage{xcolor}
\usepackage{nicefrac}
\usepackage{natbib}
\usepackage{float}

\newcommand{\affilone}{\affiliation[a]{Institut für Kernphysik, Department of Physics, Technische Universität Darmstadt, 64289 Darmstadt, Germany}}
\newcommand{\affiltwo}{\affiliation[b]{Helmholtz Forschungsakademie Hessen f\"ur FAIR (HFHF), GSI Helmholtzzentrum f\"ur Schwerionenforschung, Planckstr.1, 64291 Darmstadt, Germany}}
\newcommand{\affilthree}{\affiliation[c]{GSI Helmholtzzentrum f\"ur Schwerionenforschung GmbH, 64291 Darmstadt, Germany}}
\newcommand{\affilfour}{\affiliation[d]{Institut f\"ur Kernphysik, Universit\"at M\"unster, 48149 M\"unster, Germany}}
\newcommand{\affilfive}{\affiliation[e]{Fakult\"at f\"ur Physik und Astronomie, Universit\"at Heidelberg, 69117 Heidelberg, Germany}}
\newcommand{\affilsix}{\affiliation[f]{Helmholtzinstitut Jena, 07743 Jena, Germany}}
\newcommand{\affilseven}{\affiliation[g]{Institut f\"ur Kernphysik, Goethe Universit\"at Frankfurt, 60438 Frankfurt a. Main, Germany}}

\begin{document}

\affilone
\affiltwo 
\affilthree
\affilfour
\affilfive
\affilsix
\affilseven

\title{A Setup to Study Atomic State Population Dynamics and Optical Polarization at CRYRING@ESR}

\author[a, b, 1]{K.~Mohr,\note{Corresponding author.}}
\emailAdd{k.mohr@gsi.de}

\author[c, 1]{R.~S\'anchez}
\emailAdd{r.sanchez@gsi.de}

\author[a, b, 1]{W.~N\"ortersh\"auser}
\emailAdd{wnoertershaeuser@ikp.tu-darmstadt.de}

\author[c]{Z.~Andelkovic}

\author[d]{V.~Hannen}

\author[c]{E.-O.~Hanu}

\author[c]{F.~Herfurth}

\author[c]{R.~He\ss}

\author[a, b]{M.~Horst}

\author[a]{P.~Imgram}

\author[a, b]{K.~K\"onig}

\author[c]{C.~Krantz}

\author[c]{M.~Lestinsky}

\author[b, c, e]{Yu.~A.~Litvinov}

\author[c, f]{E.~B.~Menz}

\author[a]{P.~M\"uller}

\author[a]{J.~Palmes}

\author[a, b]{S.~Rausch}

\author[a]{T.~Ratajczyk}

\author[a]{L.~Renth}

\author[c]{J.~Rossbach}

\author[c]{R.~S.~Sidhu}

\author[a]{F.~Sommer}

\author[a]{J.~Spahn}

\author[c, g]{N.~Stallkamp}

\author[d]{K.~Ueberholz}

\author[c]{G.~Vorobjev}

\author[d]{C.~Weinheimer}

\author[a]{D.~Zisis}

\abstract{
    We present a recently established setup for laser spectroscopy at CRYRING@ESR at the GSI Helmholtz Centre for Heavy Ion Research. Here, laser spectroscopy can be performed on stored and cooled ion bunches and coasting beams. First spectra of $^{24,25}$Mg$^{+}$ ions are presented that were recorded by classical Doppler-limited fluorescence spectroscopy as well as $\Lambda$-spectroscopy using counter- and copropagating laser beams that are Doppler-shifted by several nm. 
}


\arxivnumber{2408.12368} 


\maketitle
\flushbottom

\section{Introduction}
\begin{figure}[t]
    \centering
    \includegraphics[width=.65\textwidth]{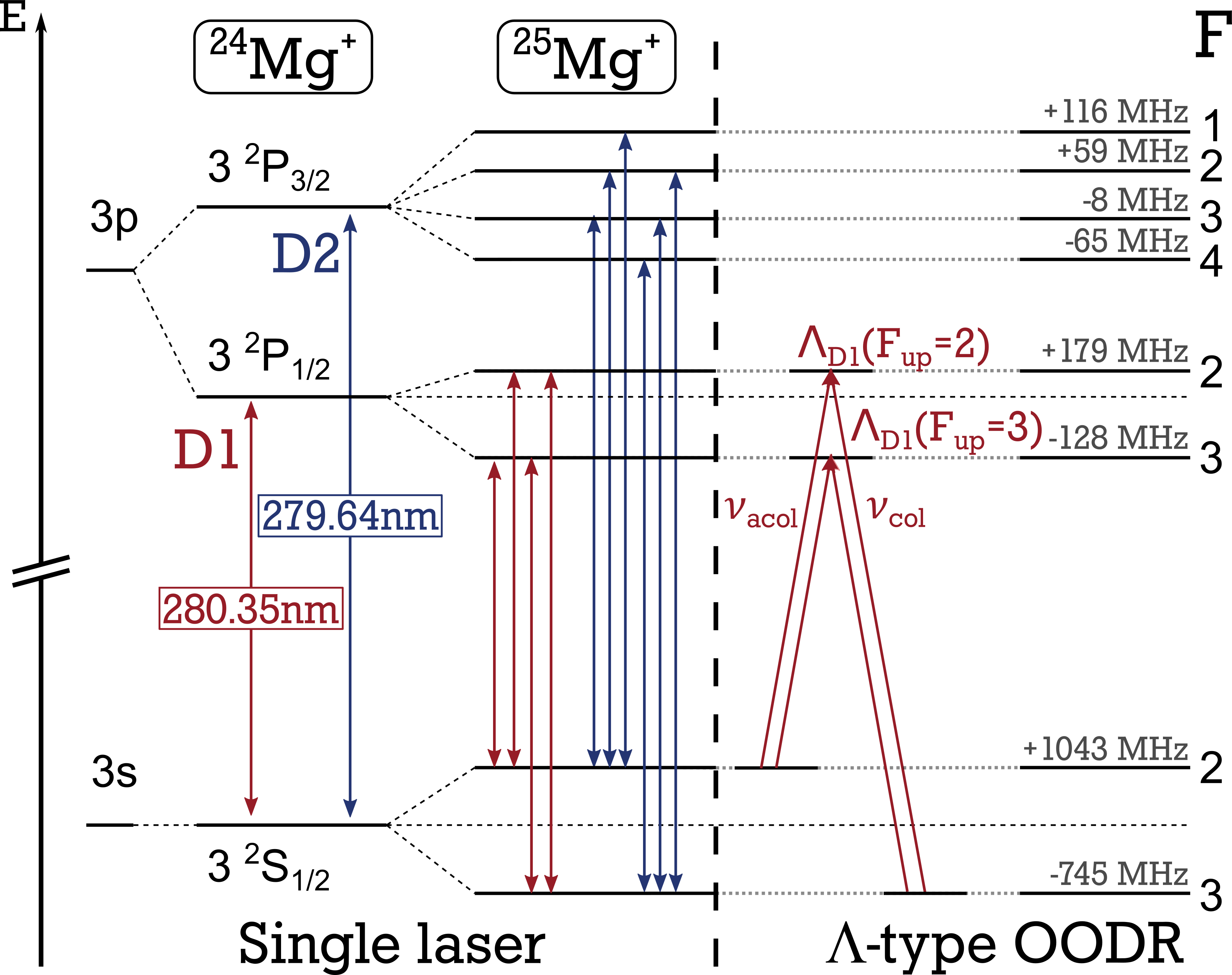}
    \caption{Level scheme of singly charged Mg-ions. Depicted are the fine structure levels in $^{24}$Mg$^+$ and the hyperfine levels in $^{25}$Mg$^+$. All possible transitions from the $^{2}$S$_{1/2}$ state that are accessible by our laser systems are indicated. Red-colored lines correspond to the D1-line with a restframe transition wavelength of $\lambda_{\mathrm{D1}} \approx \,$\SI{280.35}{nm}. Blue color indicates transitions of the D2-line with a restframe transition wavelength of $\lambda_{\mathrm{D2}} \approx \,$\SI{280.64}{nm}. If two lasers are employed, a $\mathrm{\Lambda}$-type optical-optical double resonance excitation scheme within the hyperfine structure of $^{25}$Mg$^+$ can be used to realize sub-Doppler spectral resolution.}
    \label{fig:levelScheme}
\end{figure}

Beams of spin-polarized particles, ions, and atoms are of considerable interest in fundamental and applied science \cite{Kes95, Blu99, Dap18, Mane.2005b}. Polarized electrons, muons, protons, and deuterons can be produced with high degrees of polarization from thermal to ultrarelativistic energies. The production of polarized heavy ion beams in storage rings has a huge potential for fundamental tests of the standard model \cite{Budker.2020,Bondarevskaya.2011,Labzowsky.2001}, \textit{e.g.} parity nonconservation (PNC) effects in highly-charged helium-like ions and connected with the nuclear anapole moment. Despite its potential and several suggested routes for their production, \textit{e.g.} in Refs.\cite{Prozorov.2003,Bondarevskaya.2014,Budker.2020}, such a polarized beam has not been achieved so far. Contrary, at low energies of a few \SI{10}{\keV} and for singly charged ions the interaction with circularly polarized laser beams that are superimposed with the ion beam in a weak longitudinal magnetic guiding field is a well established technique to produce polarized beams. 
They are applied to investigate the properties of short-lived nuclei \cite{Arnold.1987,Arnold.1992,Geithner.1999}, to study local magnetic fields in solid-state physics \cite{Kiefl.2003,MacFarlane.2015,Kiefl.2018}, and, most recently, for bio-NMR studies \cite{Gottberg.2014,Jancso.2017,Kowalska.2017,Szunyogh.2018}. In these applications, the ion beams are polarized by the interaction with a circularly polarized resonant laser beam that causes a redistribution of the population in the magnetic $m_J$ or $m_F$ states and, thus, introducing electronic and nuclear polarization. An observation made at the experimental storage ring ESR at the GSI Helmholtz Centre for Heavy Ion Research has given some indication for a similar behavior \cite{Nortershauser.2021} in a storage ring. This was surprising because a fast depolarization of the laser-induced population difference of the magnetic substates was expected in the large and rapidly varying magnetic fields that the ions experience during a single revolution.
To study state populations and polarization degrees of freedom, we have developed a dedicated setup at the CRYRING@ESR storage ring \cite{Abrahamsson.1993,Lestinsky.2016}. This newly installed storage ring at GSI has the advantage that it can be operated with lowly charged ions from a local ion source independently of the full GSI accelerator chain. As the ion of choice for the first experiments, Mg$^+$ has been selected because the Li$^+$ ions used at the ESR, required an ion source that provides ions in the metastable $1s2s\,^3\!S_1$ state, and Li$^+$ is a very soft beam, which makes it challenging to inject into CRYRING@ESR without prior acceleration. Since the production of the metastable state is quite inefficient, only a small fraction of the ion beam can be addressed by laser light. This is different for 
Mg$^+$ ions, which can be excited out of the ionic ground state along the $3s_{\nicefrac{1}{2}} \rightarrow 3p_{\nicefrac{1}{2},\nicefrac{3}{2}}$ transitions and can both be regarded as two-level systems for even-even isotopes like $^{24}$Mg. Additionally, the odd isotope $^{25}$Mg with a natural abundance of 10.0~\% and a nuclear spin $I = \nicefrac{5}{2}$ provides a more complex level scheme as necessary for $\mathrm{\Lambda}$-spectroscopy. Individual transitions that are addressable at CRYRING for both isotopes are shown in Fig.\,\ref{fig:levelScheme}.\\
In this article, we describe the experimental setup recently established for laser spectroscopy at CRYRING@ESR, with the goal to study optical population transfer and optical pumping, and report on first laser spectroscopic results. 

\begin{figure*}[t]
    \centering
    \includegraphics[width=.9\textwidth]{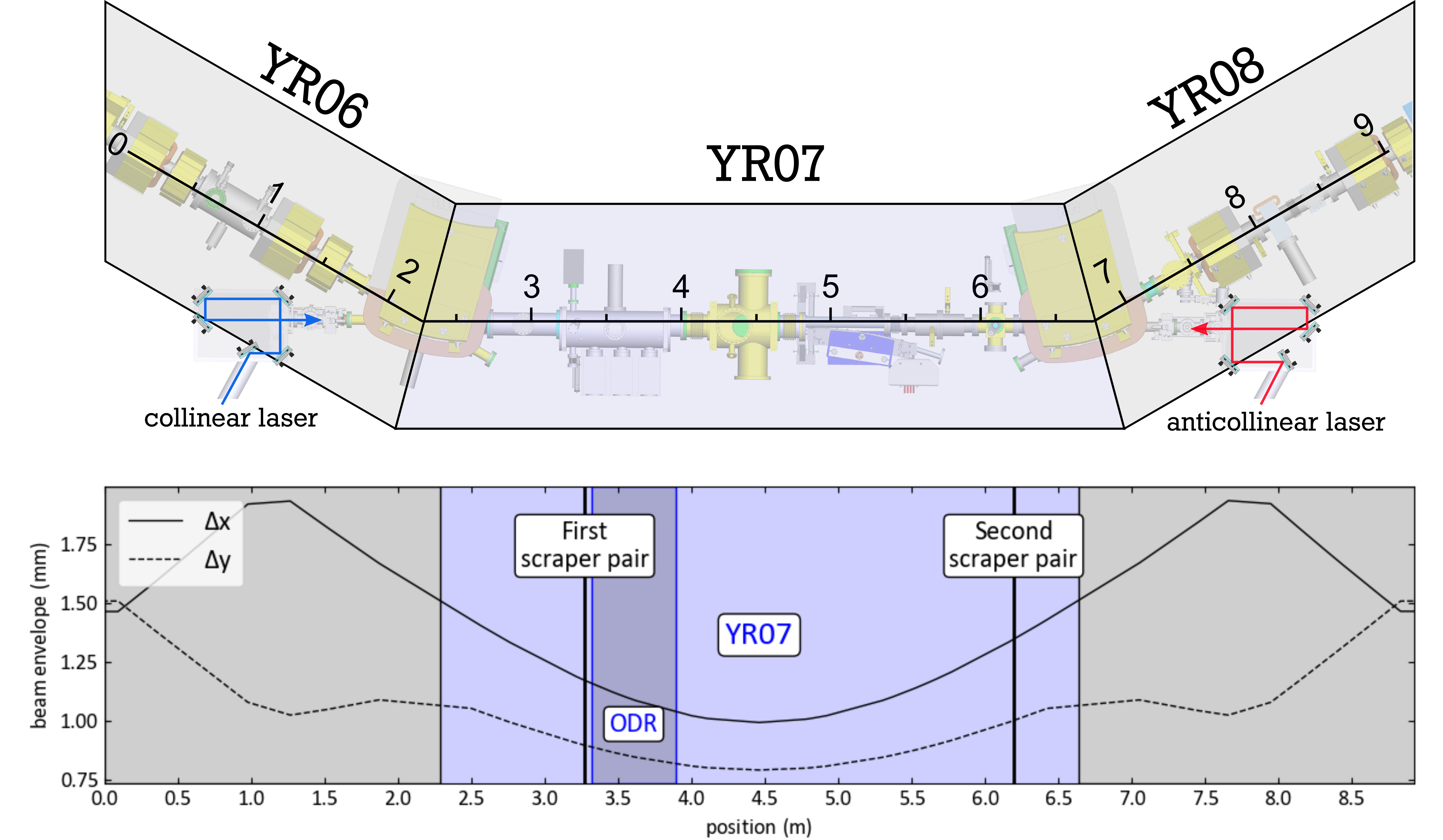}
    \caption{\textit{Top}: Overview of the sections YR06-YR08. \textit{Bottom}: Expected beam envelope as a function of the position in the ring according to the indicated scale in the upper part of the figure. The horizontal ($\Delta x$) and vertical envelope ($\Delta y$) were estimated for the emittance measurements \cite{Krantz.2021} during the beamtime in 2021 using the ion-optical lattice as presented in Gorda \textit{et al.}\cite{Gorda.2015}. The scraper pairs are indicated as black vertical lines at the positions \SI{3.27}{m} and \SI{6.20}{m}. The optical detection region (see Sec.\,\ref{subsec:ODR} and Fig.\,\ref{fig:chamber}) ranges from \SI{3.32}{m} to \SI{3.89}{m}.}
    \label{fig:YR07}
\end{figure*}

\section{Setup}
An exhaustive description of the CRYRING@ESR design has been provided in \cite{Lestinsky.2016}. Here, we will focus only on those parts that are most relevant for laser spectroscopy studies or have been explicitly designed for this task.

\subsection{Ion Beam Preparation}
Solid Magnesium is evaporated inside an oven consisting of a small ceramic vessel, heated resistively via a tungsten spiral, and equipped with a tantalum heat-shield. During the first beamtimes a Nielsen-type hot surface ion source (MINIS) \cite{Nielsen.1957} was used to create Mg$^+$ ions, which was afterwards replaced by an electron cyclotron resonance ion source (ECRIS).  
The source is located on a platform designed for bias voltages up to \SI{40}{\kV}, which is followed by a \SI{90}{\degree} sector magnet for mass separation of ion species. The subsequent radio-frequency quadrupole accelerator (RFQ) can accelerate ions up to an energy of \SI{300}{\keV} but is designed for charge-to-mass ratios of $q/m \leq \SI{3.2}{e/u}$. Thus, Mg$^+$ ions can not be accelerated but are only transported through the RFQ and injected into CRYRING@ESR at an energy of \SI{36}{\keV/q}, corresponding to \SI{1.5}{\keV/u} for $^{24}$Mg and \SI{1.44}{\keV/u} for $^{25}$Mg.
The RF system of the storage ring is then used to accelerate the ions to the designated energy. The maximum beam energies are \SI{170}{\keV/u} ($\beta=0.019$) and \SI{155}{\keV/u} ($\beta=0.0185$) for $^{24}$Mg$^+$ and $^{25}$Mg$^+$, respectively, limited by the ring's maximum magnetic rigidity of $B_\rho=\SI{1.44}{Tm}$. For convenience, the final energy is usually set to \SI{155}{\keV/u} for both isotopes. Bunching and acceleration is usually peformed on the 14$^{\mathrm{th}}$ or a higher harmonic of the RF system, resulting in 14 or more stored bunches, with subsequent rebunching or a transition to an unbunched, coasting beam.\\

\subsection{CRYRING@ESR}
The storage ring consists of 12 straight sections linked by \SI{30}{\degree} dipole bending magnets as shown in Fig.\,\ref{fig:YR07}. The sections are called 'YR$XX$'. Even numbers $XX$ refer to magnetic sections providing ion-optical focussing, odd numbered are drift sections dedicated to a specific task related to ring operation, diagnostics or experiments. The beam from the local ion source is injected into YR01, bunched and accelerated through the radio-frequency (RF) system in YR05, and electron-cooled in YR03. 
The optical detection region and the infrastructure  for laser spectroscopy experiments  is located in YR07, which is also used for beam extraction to material-science experiments. Further in-ring experiments can be installed in YR09 and beam diagnostic including the Schottky-noise pickup is mounted in YR11.

\subsection{Electron Cooler}
Beam cooling can enhance the ion lifetime in the storage ring and is essential to reduce the transversal and longitudinal emittances of the stored beam. This leads to a significant reduction of beam diameter (improved spatial laser-ion beam overlap) and velocity distribution (Doppler broadening). The electron cooler as installed at GSI has been described recently \cite{Krantz.2021} and only a brief summary is provided here. 
The adiabatic expansion technique is a peculiarity realized at the CRYRING@ESR electron cooler, which leads to low transversal beam temperatures \cite{Danared.2000}. The electron beam emerges from the electron gun at high-voltage, embedded in a strong magnetic field $B_\mathrm{g}$ of a superconducting solenoid. Two toroid-magnets are used to provide \SI{50}{\degree} bending of the electron beam for superposition with the ion beam and its subsequent extraction after a cooling section of about \SI{0.8}{m}, along which a normal-conducting solenoid provides a homogeneous magnetic flux density $B_\mathrm{d}$ along the common ion and electron beam axis in the drift section. The ratio of the magnetic flux density at the electron gun and the drift section determines the beam expansion factor $\alpha=B_\mathrm{g}/B_\mathrm{d}$, which relates the transversal electron temperature to the temperature of the electron gun ($T_{\perp}=T_{\mathrm{g}}/\alpha$). Typical expansion factors at the CRYRING@ESR electron cooler are 30--100. \\
In coasting beam operation, the electron energy in the electron cooler defines the kinetic energy of the revolving ion beam due to the thermalization between the ions and the cold electrons. Thus, it is of utmost importance for laser spectroscopy to either find a (well known) resonance transition or to extract an observed and unknown transition frequency by taking into account the large Doppler shift of the laser frequency in the ion's rest frame
\begin{equation}
  \nu_{a / c}=\nu_0 \,\gamma\,(1 \mp \beta)  
\end{equation}
for a copropagating (collinear) $\nu_c$ and counterpropagating (anticollinear) $\nu_a$ laser beam, respectively. 
In terms of the speed of light, the mean electron velocity $\beta$ -- under good cooling conditions equivalent to the ion storage velocity -- is determined from the electron acceleration voltage $U_\mathrm{eff}$ according to   
\begin{equation}
    \beta=\sqrt{1-\left(1+\frac{q U_\mathrm{eff}}{m_{\mathrm{e}} c^2}\right)^{-2}},
    \label{eq:beta}
\end{equation}
with the relativistic time-dilation factor $\gamma=1 / \sqrt{1-\beta^2}$. It should be noted that the space-charge potential $U_{\mathrm{sc}}$ of the electron beam as well as possible contact potentials $U_{\text{contact}}$ need to be considered in $U_\mathrm{eff}$. Electron beam energies for laser spectroscopy of Mg$^+$ are about \SI{90}{\eV}.
A voltage of $U_\mathrm{cooler}\lesssim \SI{102}{V}$ is applied to the electron cooler high-voltage (HV) terminal and is monitored using a high-precision HV divider ('G35'). It is a variant of the \SI{35}{\kV} precision divider designed for the KATRIN experiment \cite{Thuemmler.2009} and, together with a readout of the divider output voltage using a 8.5-digit precision digital multimeter (DMM, type Keysight 3458A), allows to determine the input voltage with an accuracy of \SI{<10}{ppm} \cite{Rest.2020}.

\subsection{Laser System and Beam Transport}

\begin{figure}[t]
    \centering
    \includegraphics[width=.6\textwidth]{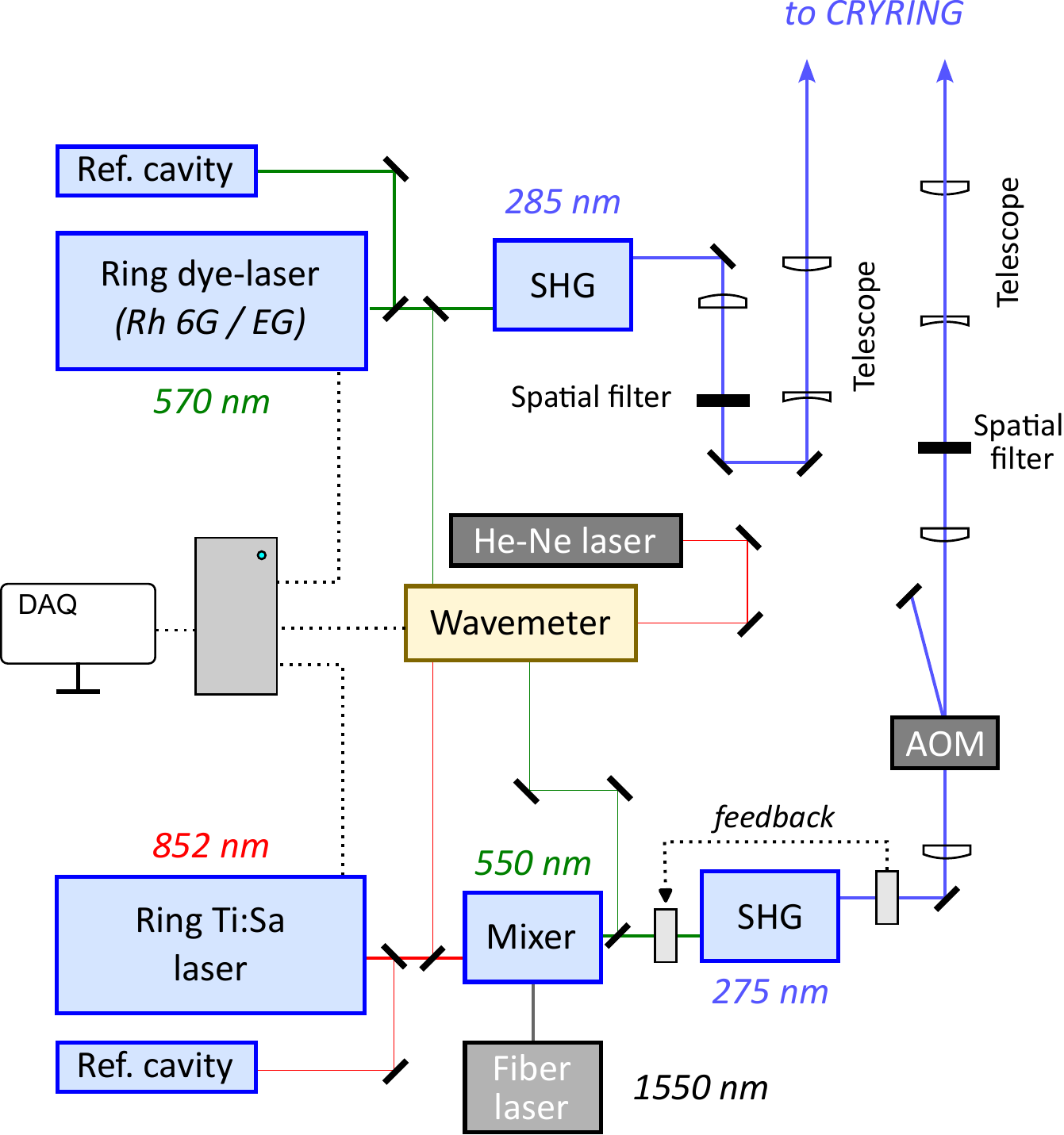}
    \caption{Overview of the laser system for the spectroscopy on Mg\textsuperscript{+}. Figure not to scale. Details see text.}
    \label{fig:lassetup}
\end{figure}

The rest-frame wavelengths of the $2s_{\nicefrac{1}{2}} \rightarrow 2p_{\nicefrac{1}{2},\nicefrac{3}{2}}$ transitions in Mg$^{+}$ are about \SI{280}{nm}. At $\beta=0.019$ the wavelength in the laboratory frame is shifted to \SI{285}{nm} for anticollinear and \SI{275}{nm} for collinear excitation. Both UV wavelengths are produced by second-harmonic generation in a Wavetrain{\textregistered}\ frequency doubler using beta-barium-borate (BBO) crystals. 
The fundamental for the anticollinear excitation (\SI{570}{nm}) is generated with a single-mode continuous-wave (cw) ring dye-laser (Matisse DS\textregistered, Sirah Lasertechnik) operated with a \SI{0.75}{g/l} Rhodamine-6G solution in ethylene glycol. It is pumped by 5--\SI{8}{\watt} of a frequency-doubled cw multimode Nd:YVO$_4$ laser (Millennia\textregistered\ eV 20, Spectra Physics). The fundamental wavelength for collinear excitation is produced by sum-frequency mixing of the \SI{852}{nm} output of a single-mode cw ring titanium-sapphire (Ti:Sa, Matisse TS\textregistered) laser and an ultra-stable fiber laser locked at a wavelength of \SI{1550.12}{nm} in a periodically poled crystal (Mixtrain\textregistered). The Ti:Sa is pumped by a second cw Millennia\textregistered\ eV laser with a power of \SI{15}{\watt}. The frequencies of both ring lasers are stabilized to a precision wavelength meter (WSU10, High Finesse), and their set-wavelengths can be controlled by the data acquisition system (DAQ). The frequency of the light obtained from mixing with the fiber laser is directly measured with the WSU10, such that the Ti:Sa frequency is adapted to accommodate also drifts of the fiber laser.

A laser power controller from Brockton Electrooptics is installed between the Mixtrain and the Wavetrain to compensate for laser power fluctuations and to compensate the power fluctuations of the Mixtrain when scanning the Ti:Sa laser frequency. This laser beam is afterwards chopped by an acousto-optical modulator (AOM) operated at a frequency of $f_{\mathrm{AOM}}=\SI{200}{MHz}$ to realize pump-and-probe experiments. Both UV-laser beams (275 nm and 285 nm) are independently cleaned using a spatial filter and then collimated using a telescope. Afterwards, both laser beams are sent to CRYRING@ESR using 2-inch optical mirrors. The settings of the optical telescopes are chosen such that the laser beams at target (section YR07) have a diameter of 2.5 mm (FWHM). An active laser beam stabilization (COMPACT\textregistered, MRC-Systems) is used for each UV-laser beam to stabilize the laser beam position at the target.

\subsection{Interaction Region and Optical Detection}
\label{subsec:ODR}
\begin{figure}[b]
    \centering
    \includegraphics[width=56mm]{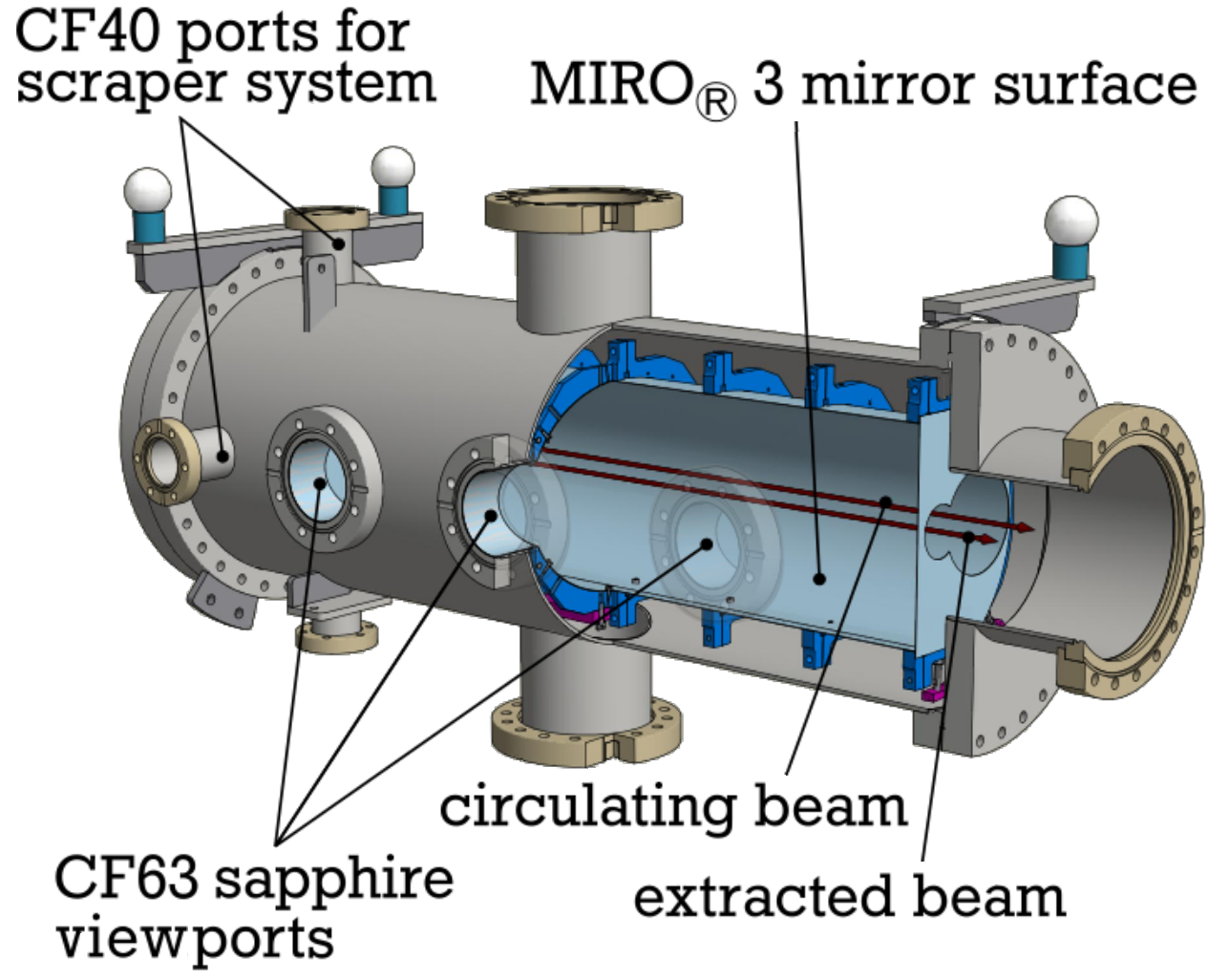}
    \includegraphics[width=28mm]{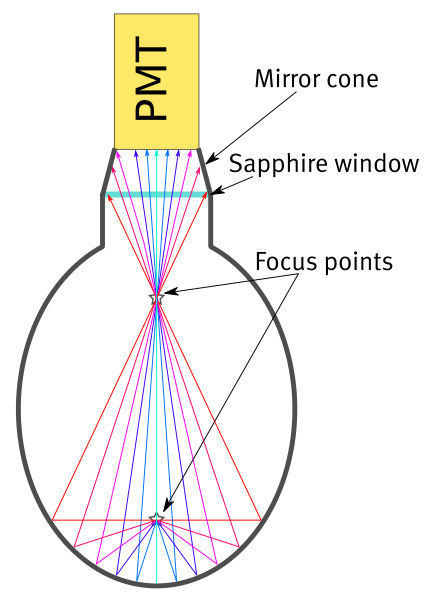}
    \caption{\textit{Left}: CAD drawing of the fluorescence detection chamber. \textit{Right}: rotated view of the elliptical cross section of the mirror chamber with exemplary photon tracks emitted by the ion beam located in the lower focal point of the ellipse.}
    \label{fig:chamber}
\end{figure}

The optical detection region for the laser spectroscopy setup at CRYRING@ESR is located at section YR07 and mounted inside a CF250 chamber having a total length of \SI{720}{\mm} (see figure~\ref{fig:chamber}, left panel). 
Seven stainless-steel ribs support an elliptical mirror system, made from MIRO\textregistered--3 aluminum sheets~\cite{miro3.2024} providing a high reflectivity down to UV wavelengths. Since this ring section is also used for slow and fast beam extraction, the end plates of the mirror system are designed with two holes providing sufficient space for the large injected beam before cooling and the extracted beam. The diameters of the holes are \SI{80}{\mm} and \SI{40}{\mm}, respectively. Several clamping screws are used to fix the support frame inside the vacuum chamber. The mirror system is divided into three identical segments of \SI{190}{\mm} length, each of them located in front of a CF63 viewport equipped with a UV-transparent sapphire window. Off-centered CF250 to CF100 adapter-flanges on both sides of the vacuum chamber ensure that the focal point of the elliptical mirror system coincides with the propagation axis of the circulating ion beam. Hence, fluorescence photons originating from the ions are collected efficiently (see figure~\ref{fig:chamber}, right panel). Photons originating from other places have a lower detection efficiency, enhancing the signal-to-noise ratio.
The whole setup can be baked at \SI{300}{\degreeCelsius} for improved vacuum conditions. \\
Outside the vacuum, each viewport is coupled to a cooled PMT housing (type FACT50, ET Enterprises Ltd.), which can be equipped with 2" photomultiplier tubes (PMTs) for single photon counting. Two sets of PMTs are available, one for the UV region (type 9235QA, ET Enterprises Ltd.) and one for longer wavelengths up to \SI{900}{\nm} (type 9658A, ET Enterprises Ltd.).
PMT signals are amplified and then digitized using a fast amplifier and a constant fraction discriminator (CFD) before being fed into the field programmable gate array (FPGA)-based data acquisition system (see below), which is a modified TILDA  \cite{Kaufmann.2019,Kanellakopoulos.2020} version tailored for laser spectroscopy at storage rings.\\
\subsection{Scraping system and spatial beam overlap}
\label{sec:scraping}
\begin{figure*}[t]
    \centering
    \includegraphics[width=.5\textwidth]{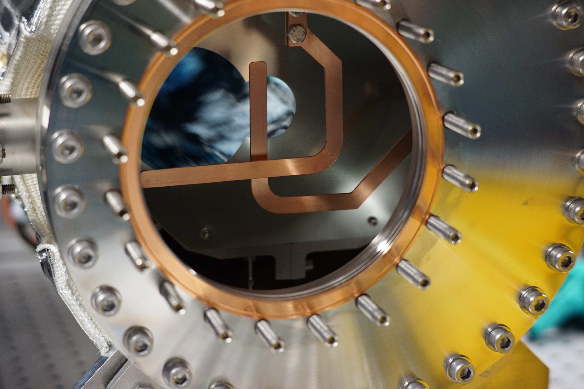}
    \hfill
    \includegraphics[width=.45\textwidth]{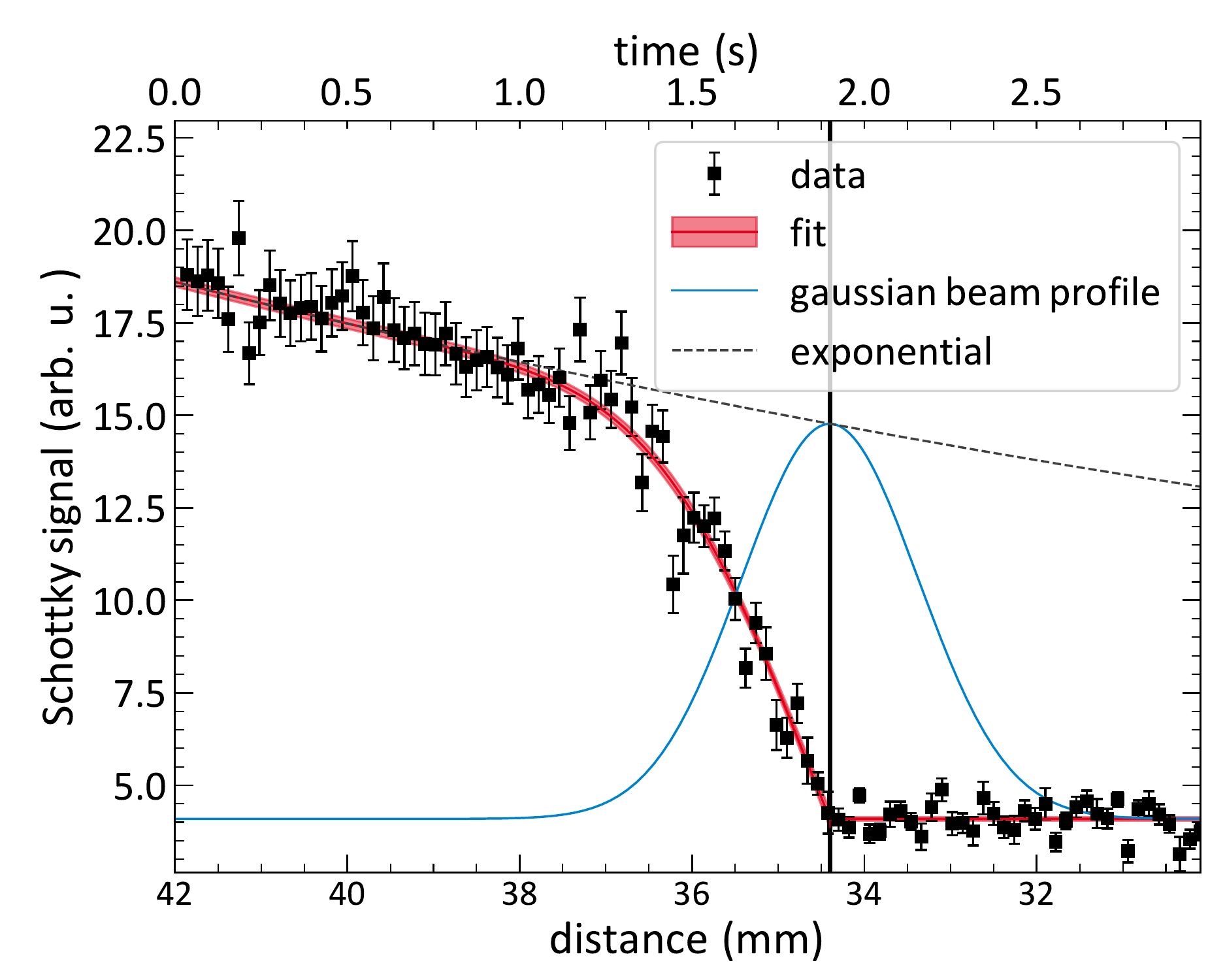}
    \caption{\textit{Left}: Picture of the scraper system mounted \SI{5}{cm} upstream the mirror chamber. \textit{Right}: Typical measurement of the pickup signal amplitude ($\blacksquare$) during a scraper measurement. While the scraper is travelling horizontally from the outside (\SI{42}{mm}) to the inside (\SI{30}{mm}), the signal decreases exponentially (dashed line) as a function of time until the scraper cuts the ion beam. From here, the signal drops more rapidly until it equals the noise level. Since the betatron oscillations are very fast ($\approx$\SI{50}{kHz})
    compared to the speed of the scraper blades ($\approx\SI{4}{mm/s}$) this position corresponds to the center of the ion beam. Equation\,(\ref{eq:beamProfile}) was fitted to the data (red curve) and can be used to infer the horizontal Gaussian beam profile (blue curve).}
    \label{fig:ScraperAndMeasurement}
\end{figure*}
In front of the mirror system one of two scraper pairs is installed, a second one \SI{2.925}{m} downstream. Each of the hook-shaped copper scraper blades can be driven either in horizontal or in vertical direction and provide sufficient space for the uncooled beam during injection, when driven to the outer or inner end-point. This allows the beam to be scraped with the \SI{10}{\mm} wide scraper blade from both sides. By simultaneously measuring the stored beam intensity, using the signal amplitude induced in a pickup electrode as a proxy, the ion beam axis can be determined in both horizontal and vertical directions. 
The scrapers are mounted on linear translators (Allectra, type DN40CF compact linear translator 100~mm/150~mm travel) with a pitch \SI{10}{mm/rev} and driven by 5-phase stepper motors. In combination with an angular resolution of \SI{1.44}{\degree} per step and a maximum step rate of \SI{1000}{Hz}, the achievable velocity of the stepper motors is $\approx\SI{4}{mm/s}$. This is sufficient to determine the ion beam position for beam lifetimes of at least a few seconds. The stepper motors are controlled via power drive cases (PDC), which are integrated into the FAIR (Facility for Antiproton and Ion Research in Europe) control system via a Stepper Motor Control Unit (microIOC). The scraper is moved into the beam from both directions to determine the beam position while observing the stored beam intensity on the pickup.
The position of the scraper can be measured in two different ways. A film resistor is used to determine an absolute position. For fast measurements of the relative position, the individual steps of the stepper motor are counted since the film resistor is limited to \SI{10}{Hz}.\\
A typical measurement of the ion beam position is depicted in Fig.\,\ref{fig:ScraperAndMeasurement}. Here, the signal-amplitude of the pickup signal is plotted against the scraper position (black points). During this measurement the scraper was driven horizontally from the starting position at \SI{42}{mm} to the end position \SI{30}{mm} counting the steps of the stepper motor. Since the lifetime of the ion beam is finite, the signal amplitude  decreases right from the start while the scraper is still travelling towards the beam's edge without disturbing it. An exponential function can express this behavior and is shown as a dashed line in Fig.\,\ref{fig:ScraperAndMeasurement}. As soon as the scraper blade interacts with the ion beam, the signal decreases more rapidly. Due to the betatron oscillations, the ion beam is completely lost once the scraper reaches the ion beam center.
To extract the ion beam profile from the data, the procedure presented for the scraper measurement at the cryogenic storage ring (CSR) \cite{Grieser.2010} was adapted. The blue curve represents a fit of the function
\begin{equation}
    I\left(x_{\mathrm{s}}\right)= \begin{cases}I_0\, \mathrm{e}^{-\frac{x_{\mathrm{s}}-x_0}{v_{\mathrm{s}} \cdot \tau}} \operatorname{Erf}\left(\frac{x_{\mathrm{s}}-x_{\mathrm{ion}}}{\sqrt{2 \pi} \sigma}\right) & x_{\mathrm{s}}-x_{\mathrm{ion}} \geq 0 \\ 0 & x_{\mathrm{s}}-x_{\mathrm{ion}}<0\end{cases} \label{eq:beamProfile}
\end{equation}
to the recorded data points, where $x_0$ and $v_{\mathrm{s}}$ are the (known) start position and drive velocity of the scraper, respectively, $x_{\text{ion}}$ the unknown center of the ion beam, \textit{i.e.}, the point at which the ion current vanishes, and $\tau$ is the ion lifetime in the storage ring. Assuming a Gaussian ion beam profile, the intensity distribution can also be modeled as shown in blue in Fig.\,\ref{fig:ScraperAndMeasurement}.  

Utilizing the visual control of the laser beam position with respect to the scraper positions allows to superimpose the ion and the laser beam with an estimated maximum displacement of $\Delta x=\Delta y = \SI{2}{mm}$ between both beams for the horizontal and vertical direction. In the worst case, the displacements at the two scraper-positions have opposite signs, resulting in a maximum angular misalignment between laser and ion beam of about \SI{2}{mrad}.

\subsection{Data Acquisition}
Laser spectroscopy at a storage ring requires the reliable combination of laser and accelerator control and readout with high timing stability. 
The data acquisition (DAQ) system must record all relevant parameters of the CRYRING@ESR operation, control the laser frequency, and collect information on the photomultiplier events. The DAQ needs to be synchronized to the operation of the storage ring. Therefore, the triggers that control the individual steps of the ion beam preparation, which are
\begin{itemize}
    \item ion creation and injection,
    \item acceleration up to the designated energy,
    \item electron cooling 
    \item measurement gate (with continuous electron cooling),
    \item extraction of the beam,
\end{itemize}
are fed into the DAQ. 
The measurement gate defines the fraction of the storage ring cycle in which laser spectroscopy can be performed. It starts when ions have reached their equilibrium temperature through electron cooling and lasts as long as a sufficient number of ions is stored in the ring. We note here that in other parts of the ion beam preparation cycle, laser spectroscopy can serve as a tool to obtain information about the ion dynamics in the storage ring, but this is beyond the scope of this work. A brief description of the implementation of the most important parameters to perform and analyze laser spectroscopy is provided in the next paragraphs.

Light ($Z \gtrsim 3$) singly charged slow ions have short lifetimes of a few seconds in CRYRING@ESR, and the corresponding fraction of the complete accelerator cycle for laser spectroscopy is relatively small. To improve the duty cycle, accelerator cycles of 7--\SI{10}{s} are typically applied of which 1--\SI{3}{s} are used for spectroscopy. The laser frequency is typically changed once the extraction trigger is received to record a spectrum. This provides sufficient time to set the next laser frequency before the next prepared ion beam is delivered by the ring. The whole storage time is used to determine the fluorescence signal at the set laser frequency. 

Photomultiplier signals are processed in the FPGA (NI PXI-7852R) to obtain the count rate and the photon's arrival time. All arrival times are recorded relative to a fixed phase of the ion revolution. Therefore, the discriminated bunching frequency obtained from the ring RF system is divided by the harmonic number and used as the trigger input.

The bunch-pickup signal mentioned above provides ion-current information and is used to normalize the fluorescence signal (see below). The bunch-pickup system consists of four conductive plates in YR11, also used for Schottky analysis of coasting, highly-charged ion beams.
Induced voltages are fast-amplified and subsequently Fourier-transformed by a real-time spectrum analyzer (RSA, type Tektronix N9020B MXA Signal Analyzer). For bunched beam operation, the dominating frequency component is given by  the bunching frequency, \textit{i.e.}, the revolution frequency (for Mg$^+$ under the usual conditions about \SI{100}{\kHz}) times the number of bunches. The signal of this dominant frequency component, taken from the analog output of the RSA, is fed into a voltage-to-frequency converter with a conversion ratio of \SI{1}{MHz/V}. The resulting frequency is recorded by the real-time DAQ to provide a relative ion current measurement.

The platform voltage and the electron current at the electron cooler are continuously published via the open-source event streaming system Apache Kafka\textregistered, which is integrated into the CRYRING@ESR data acquisition. From here, the information can be read from any device and stored for analysis of the electron velocity $\beta$ according to Eq.\,(\ref{eq:beta}).

\section{First Results}

\begin{figure*}[t]
    \centering
    \includegraphics[width=\textwidth]{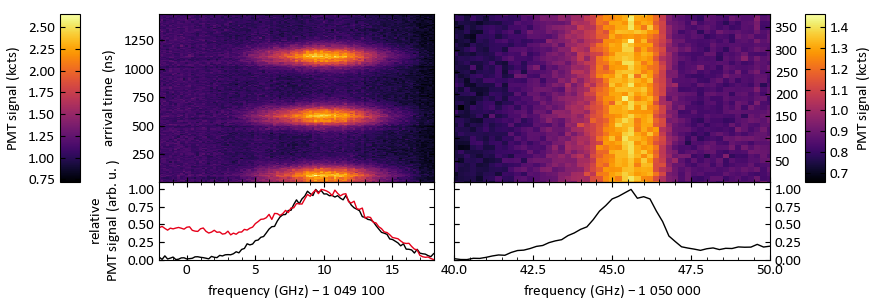}
    \caption{\textit{Left}: Spectrum of the $4s_{\nicefrac{1}{2}} \rightarrow 4p_{\nicefrac{1}{2}}$ transition (D1 line) in $^{24}$Mg$^{+}$ obtained with bunched-beam operation in anticollinear geometry and without laser power controller. The density plot depicts the photomultiplier (PMT) signal as a function of the arrival time relative to a certain phase of the RF signal ($y$) and the laser frequency ($x$). Three ion bunches are visible in the left subfigure, taken with a bunched ion beam. The projection in the lower frame represents the relative fluorescence signal as a function of the laser frequency. The raw signal (red curve) is normalized using the laser background signal between the ion bunches resulting in the Gaussian-like resonance (black curve). \textit{Right}: Spectrum of the D1-line in $^{24}$Mg$^{+}$ taken with a coasting beam. The fluorescence signal is continuous in the time domain. Normalization of laser and ion-beam intensity is more challenging in this case and is described in detail in the text. 
    }
    \label{fig:BunchedVsCoasting}
\end{figure*}

We present spectra of Mg$^+$ ions obtained at CRYRING@ESR with the above-described setup under several experimental conditions. 
\subsection{Doppler-limited Spectroscopy}
\begin{figure}[t]
    \centering
    \includegraphics[width=.65\textwidth]{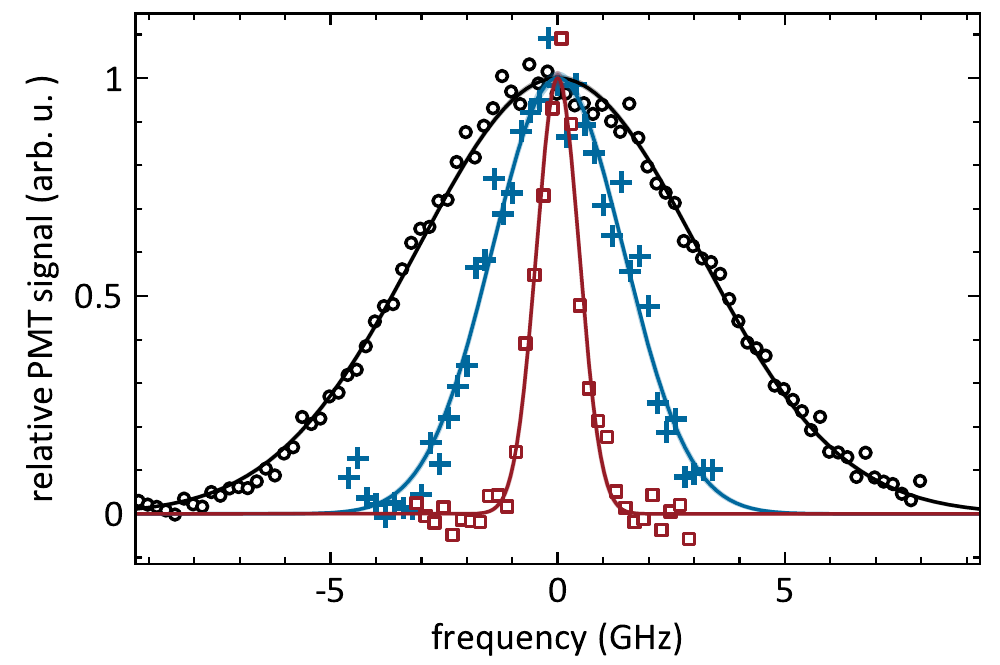}
    \caption{Comparison of normalized spectra of the D1-line in $^{24}$Mg$^{+}$ obtained with a bunched beam. The black data points ($\circ$) were recorded during the first beamtime in 2019 without electron cooler and are the same data as in Fig.\,\ref{fig:BunchedVsCoasting}a. The full-width-at-half-maximum (FWHM) of the Gaussian line profile is determined to \SI{7.29}{GHz}. During the measurement taken in the 2020 beamtime  (\textcolor{blue}{$+$}), electron cooling was not optimal due to a slight misalignment between the cooler's guiding magnetic field orientation and the electron-gun direction. This lead to insufficient cooling, but still a clearly reduced Doppler width. Finally, with proper electron cooler settings, a FWHM of \SI{1.11}{GHz} was achieved in 2021 (\textcolor{red}{$\square$}), a factor of 7 reduced compared to the resonance taken without electron cooling.}
    \label{fig:CoolingComparison}
\end{figure}

Figure\,\ref{fig:BunchedVsCoasting} shows resonance signals of a bunched (a) and a coasting beam (b). The laser was scanned across the resonance position, and the photomultiplier signals were recorded using the laser frequency. Due to the short lifetime of the ions in the ring of only a few seconds, the spectra are taken with a single fixed laser frequency for each ring filling. The laser is tuned to the next frequency, while the ions of the next injection are prepared and cooled. The photon's arrival time related to an arbitrary but fixed phase of the revolution frequency was also recorded. The color-coded signal intensity in the upper part of the figure shows the result, where the $x$-axis is the laser frequency while the $y$-axis represents the photon arrival time. The 18$^\text{th}$ harmonic of the revolution frequency was applied at the RF-cavity. Three out of the 18 revolving bunches are visible in the fluorescence spectrum. 
The lower frame shows the extracted resonance signal before (red) and after (black) normalization. The red one is obtained by simply projecting the number of counts to the $x$-axis. However, the intervals in which ions are not present in the detection region were disregarded to reduce the laser-induced background. The obtained signal was recorded in anticollinear geometry without the laser power controller and exhibits a clear asymmetry with a falling baseline. It turned out that this is largely caused by laser intensity fluctuations and shot-to-shot ion intensity variations while recording the spectra. Normalization to the laser intensity was performed by integrating the dark counts in the periods where no ion bunch is present at the optical detection, \textit{i.e.}, those regions previously not included in the summation. The ion beam intensity is obtained from the bunch pickup signal, as described in Sect.\,\ref{sec:scraping}, and can also be used in the normalization procedure (not applied here). The normalization leads to a reasonable Gaussian-like structure as shown by the black trace in Fig.\,\ref{fig:BunchedVsCoasting}a. 
Lineshapes with bunched beams from different beamtimes are presented in Fig.\,\ref{fig:CoolingComparison}. They are well described by Gaussian lineshapes with decreasing linewidth (FWHM): In the first beamtime in 2019 ($\circ$) the electron cooler was not operated (\SI{7.3}{GHz)}. In 2020 (\textcolor{blue}{+}) it was not fully optimized (\SI{4.46}{GHz}) and, finally, in 2021 (\textcolor{red}{$\square$}) optimal conditions were achieved (\SI{1.11}{GHz}), providing a seven-fold improvement in linewidth compared to operation without cooling. 

As bunching introduces some heating due to driven synchrotron oscillations, the limit is reached if the heating rate equals the cooling rate. It can, therefore, be expected that the beam temperature can be further decreased if the bunching is turned off, \textit{i.e.}, coasting beam operation is used. 
A spectrum taken with a coasting beam (December 2023) is shown in Fig.\,\ref{fig:BunchedVsCoasting}b. No bunches are observed in this case. Ion arrival and fluorescence light are equally distributed over the revolution time. The lineshape has a tail to lower frequencies, which correspond to higher velocities of the ions in counterpropagating geometry. 
Without RF-bunching, ion-current normalization is not possible due to the absence of a strong coherent pickup signal. The Schottky noise of the coasting beam is undetectable with our present pickup system at the low charge state and revolution frequency of the Mg$^+$ beam.
Laser intensity normalization using intrinsic information is also not possible since periods without ion beam in the detector region are not available. The origin of the tail is not yet fully understood but there are indications that it might be related to a not fully converged cooling process when the spectroscopy gate was started. We also note that the linewidth of \SI{3.71}{GHz} is larger than the one observed in bunched-beam mode in the 2021 beamtime (\SI{1.11}{GHz}) and rather comparable to the linewidth of \SI{4.46}{GHz} observed in 2020. This is in conflict with the expectation of a lower linewidth for a coasting beam under identical cooling conditions and indicates that the electron cooling was not fully optimized. Data analysis and simulations of these spectra are currently performed.

\subsection{Sub-Doppler $\Lambda$-Spectroscopy}
\begin{figure}[t]
    \centering
    \includegraphics[width=.45\textwidth]{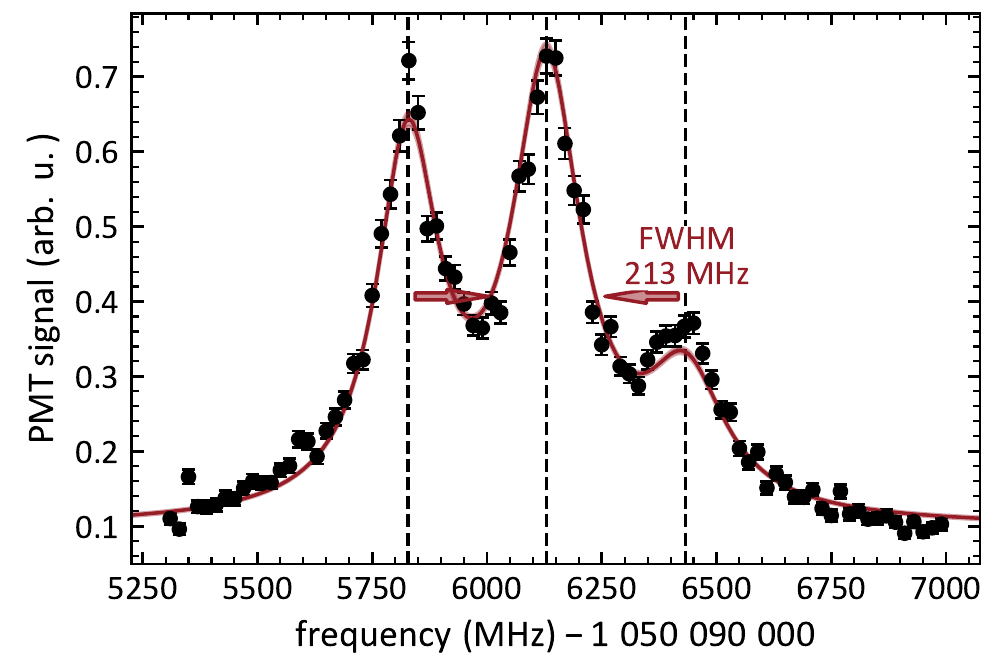}
    \includegraphics[width=.35\textwidth]{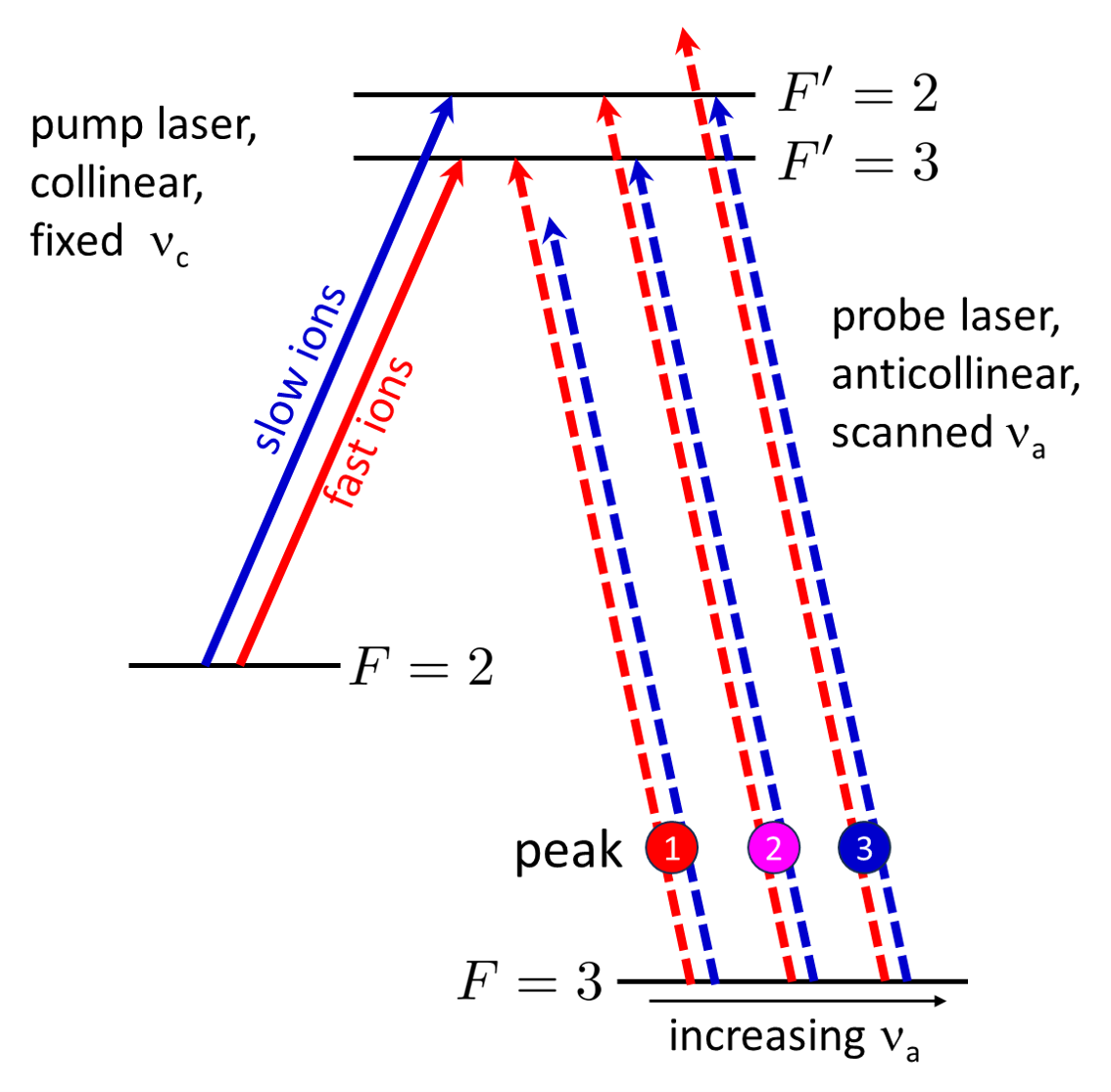}
    \caption{\textit{Top}: Spectrum of the D1-line in $^{25}$Mg$^{+}$ using $\mathrm{\Lambda}$-type optical-optical double resonance technique with a collinear laser frequency fixed at \SI{1\,088\,936\,500}{MHz} and a scanning anticollinear laser. The resulting FWHM of \SI{213}{MHz} is more than twice as large as expected for a natural linewidth of $\Gamma=\SI{41.3}{MHz}$. The three dashed lines indicate the positions of the individual hyperfine lines.
    \textit{Bottom}: Explanation of the origin of the three peaks: 
    The pump laser (solid line) is simultaneously in resonance with two velocity groups within the Doppler distribution of the ion beam. Fast ions are driven to the lower $3p_{\nicefrac{1}{2}} \, F^\prime=3$ hyperfine state, slower ions to the upper $F^\prime=2$ state. The groups of red and blue dashed lines represent the laser frequency of the anticollinear probe laser in the rest frame of the fast and slow velocity groups, respectively, at the laser frequencies corresponding to the three peaks observable in the spectrum.  
    }
    \label{fig:LambdaSpectrum}
\end{figure}

Finally, we present a first spectrum of $^{25}$Mg$^{+}$ taken with $\mathrm{\Lambda}$-spectroscopy in the $3s_{\nicefrac{1}{2}} \rightarrow 3p_{\nicefrac{1}{2}}$  transition. The more intense copropagating laser is fixed to the $F=2 \rightarrow F^\prime=3$  transition close to the maximum of the Doppler-broadened resonance, while the frequency of the weaker counterpropagating laser is scanned across the transitions starting on the  $F=3$ hyperfine level. The principle of $\mathrm{\Lambda}$-spectroscopy is as follows: as long as both laser beams are operating on different velocity classes of the revolving ions, the respective ions will be quickly pumped into the other (dark) hyperfine level of the electronic ground state from where no excitation can occur anymore. A repeated transfer between the two ground-state hyperfine levels and, thus, enhanced fluorescence becomes possible only when the probe laser frequency simultaneously addresses those ions that are pumped by the second laser. 
While the spectrum of the even isotope in Fig.\,\ref{fig:BunchedVsCoasting}b represents the total Doppler width of the ion velocity distribution, $\mathrm{\Lambda}$-spectroscopy provides, in principle, spectra that exhibit a resolution given by twice the (homogeneously broadened) natural linewidth of the transition. The factor of two arises from the fact that the pumped fraction represents a Lorentzian distribution in velocity space with width corresponding to the natural linewidth, which is then convoluted with the probing line profile, which has approximately the same width and shape. The natural linewidth of the D1 line of Mg$^+$ is \SI{41.3}{MHz}, which might be slightly power broadened under our experimental conditions. The lineshape in Fig.\,\ref{fig:LambdaSpectrum} is well represented by a Lorentzian shape with a FWHM of \SI{213}{\MHz}, which is slightly more than twice the expected value. Additional broadening can be caused by velocity changing collisions, \textit{e.g.}, in the cooler, as it was also observed in the $\mathrm{\Lambda}$-spectroscopy of Li$^{+}$ ions in the ESR, where the natural linewidth was of the order of \SI{5}{\MHz} but an experimental linewidth of about \SI{100}{\MHz} \cite{Botermann.2014} was observed.  

Three peaks are visible in the spectrum even though the $3p_{\nicefrac{1}{2}}$ level has only two hyperfine states ($F^\prime=2,3$). It turns out that the distances from the central peak to both, the left and the right peak, are after correction for the relativistic Doppler-Shift in reasonable agreement with the well-known hyperfine structure splitting in the $3p_{\nicefrac{1}{2}}$ level $\Delta\nu_{3p-\mathrm{hfs}}=\SI{308.61(24)}{MHz}$ \cite{Hao.2023}. The third peak's appearance is due to the relatively broad velocity distribution that covers both hyperfine states of the $3p_{\nicefrac{1}{2}}$ level within the Doppler width. Therefore, both lasers are in resonance with the upper ($F^\prime=2$) state in one velocity class and the lower ($F^\prime=3$) state in another. This is visualized in the lower part of Fig.\,\ref{fig:LambdaSpectrum}. There are three possibilities for cooperation between the velocity groups. The left peak in Fig.\,\ref{fig:LambdaSpectrum} corresponds to the case where the fast-ion fraction of the anticollinear probe laser excites the $F=3 \rightarrow F'=3$ transition while the same (fast) ions interact simultaneously with the collinear pump laser on the $F=2 \rightarrow F'=3$. Hence, they share the same upper state to pump the population between the two ground-state hyperfine levels. If the frequency of the probe laser is now increased by $\Delta\nu_{2p-\mathrm{hfs}}$, the fast fraction of the probed ions coincides with the slow-ion fraction of the pump laser, while simultaneously, the slow fraction of the probe laser coincides with the fast-ion fraction driven by the pump laser. At this point, both transitions and both velocity groups contribute to the fluorescence, which explains the large intensity of the central peak. Increasing the anticollinear probe laser frequency again by $\Delta\nu_{2p-\mathrm{hfs}}$, the $F=3 \rightarrow F^\prime=2$ transition of the slow ion fraction is addressed while the same ions are pumped back by the collinear laser on the $F=2 \rightarrow F^\prime=2$ transition. The situation resembles somehow the crossover resonances in standard saturation spectroscopy but is nevertheless different since the resonances do not appear halfway between the "real" saturation signals but have the full hyperfine splitting. 

\section{Conclusion and Outlook}
We have established a setup to perform collinear laser spectroscopy at CRYRING@ESR. Collinear and anticollinear excitation was performed and simultaneous operation in both directions has been demonstrated for using $\mathrm{\Lambda}$-spectroscopy. Lineshapes have been studied with and without electron cooling and for both modes of operation, \textit{i.e.}, bunched-beam and coasting-beam operation. Linewidths between \SI{1}{\GHz} and \SI{7}{\GHz} have been observed. They were further reduced to about \SI{200}{\MHz} with the application of $\mathrm{\Lambda}$-spectroscopy. With these installations, we have established laser spectroscopy as a versatile tool at CRYRING@ESR and will use it, \textit{e.g.}, to continue studying the application of optical pumping in magnetic storage rings. At CRYRING@ESR we plan to exploit excitation schemes, which ensure that the same velocity class is addressed over many revolutions. A vanishing or at least a reduced fluorescence signal would be observable for a possible polarization build-up. However, such a signal has to be carefully distinguished from other side effects like synchrotron (only during bunched beam operation) or betatron motion, which vary the resonance condition with time. The presented $\Lambda$-scheme applied to $^{25}$Mg$^+$ can be used to investigate state population, and polarization dynamics when the laser beams are circularly polarized. Restricting the interactions to every $n$-th revolution will provide insight into the temporal evolution and quantify the competitive processes due to the ion dynamics. In a further step the excitation region can be separated from the optical detection region, i.e., at the electron cooler section. This would have the advantage of providing a magnetic guide field and would allow using pulsed laser systems with a spectral width that ideally matches the width of the ion velocity distribution, resulting in optical pumping for all velocity classes simultaneously.

\ \\

\section{Acknowledgement}
The research presented here is a result of a R\&D project experiments E148 and  G-22-00058 at CRYRING@ESR in the frame of FAIR Phase-0 supported by the GSI Helmholtz Centre for Heavy Ion Research in Darmstadt (Germany) and the German Minsitry for Education and research (BMBF) under contracts 05P21RDFA1 and 05P21PMFA1.


\bibliographystyle{bst/JHEP.bst}
\bibliography{references}

\end{document}